# Large Language Model Counterarguments in Older Adults: Cognitive Offloading or Susceptibility to Moral Persuasion?

Kou Tamura, Sayaka Ishibashi, Ayana Goma, Kenta Yamamoto, Kouhei Masumoto

Kobe University

**Corresponding author**

Correspondence concerning this article should be addressed to Kouhei Masumoto, Graduate School of Human Development and Environment, Kobe University, 3-11 Tsurukabuto, Nada-ku, Kobe, Hyogo 657-8501, Japan. TEL: +81-78-803-7892. E-mail: masumoto@people.kobe-u.ac.jp

**Abstract**

This study examined whether counterarguments generated by large language models (LLMs) influence the moral judgments of younger and older adults, and whether these effects vary by dilemma type, cognitive functioning, trust in AI, and prior LLM experience. Using the switch and footbridge trolley dilemmas, 130 participants (56 younger adults and 74 older adults) were presented with ChatGPT-generated counterarguments that opposed their initial judgments. More than 30% of participants reversed their judgments in both dilemmas (32.31% in the switch dilemma and 36.92% in the footbridge dilemma). Older adults tended to be more likely than younger adults to reverse their judgments and showed a significantly greater degree of judgment change in the switch dilemma. In the emotionally aversive footbridge dilemma, older adults with lower cognitive functioning were significantly more likely to align with the LLM-generated counterargument. General trust in AI and prior LLM experience did not predict judgment reversal, whereas lower initial confidence and higher perceived task difficulty were associated with greater susceptibility to LLM influence. These findings suggest that LLMs may support cognitive offloading but increase susceptibility among individuals with limited cognitive resources. The ecological generalizability of these findings to everyday dilemma situations remains to be examined in future research.

## 1. Introduction

Large language models (LLMs) have been used as tools of persuasive communication that influence people's thoughts, choices, and behaviors in domains such as political communication and marketing (Hölbling et al., 2025). At the same time, they have been criticized for their potential to generate and disseminate hate speech, misinformation, and political propaganda (Salvi et al., 2025). Although a meta-analysis on AI persuasiveness found no consistent difference in persuasive effectiveness between humans and LLMs, it has emphasized the need to focus more closely on individual characteristics and the psychological factors underlying persuasion (Hölbling et al., 2025). The purpose of the present study is to examine whether individuals reverse their judgments and accept LLM-generated counterarguments when the LLM presents a position opposite to their original judgment. The counterarguments were designed to identify potential problems with participants' initial judgments through rational arguments based on multiple perspectives, rather than relying on emotional appeals or unsupported rejection. This study also investigates whether the acceptance of such counterarguments is influenced by individual attributes, such as age, as well as psychological factors, such as cognitive function and trust in AI.

Despite the growing use of LLMs as persuasive agents, it remains insufficiently understood who is susceptible to their influence, and under what conditions, especially when the issue at stake involves morally consequential judgments for which there is no objectively correct answer. This question is particularly important for older adults. In rapidly aging societies, older adults are increasingly likely to encounter AI tools (Abadir et al., 2024). Examining how cognitive functioning is associated with older adults' acceptance of LLM-generated counterarguments therefore provides important insights into the influence of LLMs on decision making. The present study makes three academic contributions. First, rather than examining advice or recommendations provided by LLMs, it focuses on situations in which an LLM directly challenges participants' own prior judgments. Second, it includes older adults and directly assesses cognitive functioning as a predictor of responsiveness to LLM-generated arguments. Third, it distinguishes subjective evaluations of one's own judgment, namely confidence and perceived difficulty, from general attitudes toward AI, namely trust in AI, and examines which of these factors predicts responsiveness to LLM persuasion. The theoretical framework and the specific hypotheses derived from it are developed below.

## 2. Theoretical Framework

In the present study, the switch dilemma and the footbridge dilemma are used as decision-making tasks without objectively correct answers to examine whether people's judgments change in response to LLM-generated counterarguments. The switch dilemma is a task in which participants judge whether it is appropriate to pull a switch to divert a trolley, thereby sacrificing one person in order to save five people who would otherwise be killed if the trolley continued on its current course. The footbridge dilemma is a task in which participants judge whether it is appropriate to push a man off a footbridge to stop the trolley and save five people. These two trolley dilemmas are equivalent in that both involve sacrificing one life to save five others. However, previous research has shown that many individuals make a utilitarian judgment in the switch dilemma, favoring the sacrifice of one person in order to save five because doing so maximizes benefits and minimizes costs. In contrast, in the footbridge dilemma, individuals tend to make a deontological judgment, showing a moral aversion to the act of pushing a person to his death (Greene et al., 2001).

One theoretical framework that explains the difference in judgments between these two dilemmas is the dual-process theory of moral judgment (e.g., Greene, 2009). According to this model, moral decision-making is shaped by two qualitatively distinct processes: an emotional and automatic process, and a controlled cognitive process that can override or modify judgments based on the former. In the switch dilemma, individuals are thought to rely more heavily on controlled cognitive processing because the judgment requires weighing the costs and benefits of sacrificing one person to save five others. In contrast, in the footbridge dilemma, the act of directly pushing a person to their death elicits a strong negative emotional response, leading individuals to rely more on emotional and automatic processing.

It is important to note, however, that the dual-process framework primarily concerns how individuals generate their initial moral judgments, rather than how such judgments are revised when individuals are confronted with persuasive counterarguments. Recent work has called for a revision of the dual-process framework, as foundational assumptions about the exclusivity of intuitive and deliberate processes remain contested (De Neys, 2023; Greene, 2023). Whether the cognitive systems implicated in initial judgment formation also govern responsiveness to external counterarguments thus remains an open empirical question. The present study therefore offers an empirical test of whether the dual-process distinction generalizes from initial judgment formation to judgment revision in the face of LLM-generated counterarguments.

## 3. The Present Study and Hypotheses

Building on the theoretical framework outlined above, we derived the following hypotheses concerning whether LLM-generated counterarguments change moral judgments and whether factors such as age, cognitive functioning, and trust in AI influence the acceptance of these counterarguments in moral judgment contexts. Because the present study was not preregistered (see Open Practices Statement), and because some of the relevant literatures support competing directional predictions, the hypothesis tests reported below should be regarded as exploratory rather than strictly confirmatory. We state each prediction explicitly, together with its theoretical grounding, so that the basis for each test is transparent.

### *3.1.Does the Persuasive Effect of LLM-Generated Counterarguments Depend on Dilemma Type?*

The present study examines the extent to which individuals revise their judgments when an LLM presents a counterargument to their original judgment. Previous research has shown that judgments in moral dilemmas can be influenced by LLM opinions (Krügel et al., 2023), and recent work suggests that people may treat LLM-generated moral explanations as credible sources of moral reasoning (Dillion et al., 2025; Landes et al., 2026). However, previous studies have typically presented LLM outputs as advice or moral evaluations, rather than as direct counterarguments to participants' own prior judgments. Thus, it remains unclear how often individuals reverse their judgments when an LLM explicitly challenges their initial position.

The present study further investigates whether this effect differs between the switch dilemma and the footbridge dilemma, which have been theorized to place different relative demands on controlled cognitive processing and emotional and automatic processing. Because moral judgments grounded in strong affective responses are often resistant to revision through deliberative reasoning (Haidt, 2001), we predicted that participants would be more likely to accept counterarguments in the switch dilemma than in the footbridge dilemma.

**H1:** Individuals will be more likely to accept LLM-generated counterarguments and change their judgments in the switch dilemma than in the footbridge dilemma.

This prediction follows from the traditional dual-process account. However, recent evidence suggests that utilitarian responses to even emotionally weighty moral dilemmas can be generated intuitively rather than necessarily through deliberate correction of an initial deontological response, indicating that both utilitarian and deontological considerations may be available as competing intuitions within moral dilemmas (Bago & De Neys, 2019). LLM-generated counterarguments may therefore activate or reinforce the competing intuition in both the switch and footbridge dilemmas. This makes the direction and magnitude of dilemma-type differences in judgment reversal difficult to specify a priori. Thus, H1 was treated as exploratory.

### *3.2 Does Age Influence the Acceptance of LLM-Generated Counterarguments?*

The present study focuses on age as an individual characteristic that may influence the acceptance of LLM-generated counterarguments. Because aging is associated with declines in long-term memory, processing speed, working memory, and executive functioning (Salthouse, 2010, 2019), older adults often experience greater difficulty with decisions that require the evaluation of complex options or probabilities, rapid judgment, information search, and novel decision making (Mather, 2006; Peters et al., 2007). Given that aging affects decision-making processes, age may be an important individual characteristic to consider when examining the influence of LLM-generated counterarguments on decision making.

For example, older adults tend to strategically avoid processing information that is not personally important or engaging in effortful thought in order to conserve cognitive resources (Hess, 2014). Accordingly, older adults may be less likely to engage in demanding cognitive strategies, such as carefully evaluating counterarguments from an LLM and acknowledging the possibility that their own initial judgment was incorrect. If so, they may be less likely than younger adults to accept LLM-generated counterarguments and revise their judgments. In addition, older adults have been shown to exhibit a greater tendency toward deontological judgment than younger adults and may be more reluctant to act in ways that conflict with moral idealism, which emphasizes adherence to socially defined norms (McNair et al., 2019). Therefore, compared with younger adults, older adults are expected to be less likely to accept LLM-generated counterarguments in the footbridge dilemma, which elicits stronger negative emotional responses.

From the perspective of cognitive offloading, however, the opposite prediction can be derived. Cognitive offloading refers to the use of external tools to reduce cognitive load, including the use of notes, reminders, and digital tools such as AI (Gerlich, 2025; Risko & Gilbert, 2016). For example, in prospective memory tasks, older adults have been shown to engage in cognitive offloading by using reminders more frequently than younger adults to reduce cognitive demands (Gilbert, 2015). In addition, it has been reported that the use of LLMs during information search in decision-making contexts can reduce the perceived difficulty of decision making among older adults (Ishibashi et al., 2025). Because cognitive offloading provides a more direct theoretical link between age-related cognitive decline and the use of LLM-generated information, we adopt this perspective as the basis for our formal prediction in the present study.

**H2**: Older adults will be more likely than younger adults to change their judgments in response to LLM-generated counterarguments, and this tendency will be especially strong among those with lower cognitive functioning.

Most prior research on interactions between LLMs and humans has relied on data from younger populations, particularly university students and participants in online surveys, whereas studies involving older adults remain extremely limited (Ishibashi et al., 2025). Moreover, to the best of the authors' knowledge, no previous study has directly assessed cognitive functioning and examined whether cognitive decline influences changes in judgment in response to information provided by LLMs.

### *3.3 What Individual Factors Influence the Acceptance of LLM-Generated Counterarguments?*

In addition to age, other individual factors may also influence the acceptance of LLM-generated counterarguments. Research on advice taking has shown that individuals are less likely to accept advice when they have greater confidence in their own initial judgment, whereas they are more likely to accept advice when the task is difficult or when they perceive the advice source as reliable (Bonaccio & Dalal, 2006; Gino & Moore, 2007; Yaniv, 2004). Accordingly, the present study examined whether initial confidence and perceived decision difficulty predict judgment reversal following LLM-generated counterarguments.

The present study also examined trust in AI and prior experience using LLMs. Trust was included because advice taking depends not only on the content of advice but also on the perceived reliability and competence of the advice source (Bonaccio & Dalal, 2006; Yaniv, 2004). In human-AI interaction, trust is often treated as a key determinant of whether users accept, rely on, or comply with algorithmic recommendations. This rationale is especially relevant in the present context, as recent work suggests that people may perceive LLM-generated moral explanations as thoughtful, trustworthy, and credible sources of moral reasoning (Dillion et al., 2025). Relatedly, research on AI-assisted moral decision-making indicates that recommendations from autonomous systems can shape moral choices and alter perceived responsibility in ethically charged contexts (Salatino et al., 2025). At the same time, people often exhibit algorithm aversion, that is, a tendency to prefer human judgment over algorithmic judgment, especially in tasks that are open to interpretation and rely on personal intuition (Castelo et al., 2019). Because moral dilemma tasks, particularly the footbridge dilemma, elicit strong negative emotions and involve inherently subjective judgments, whether general trust in AI facilitates acceptance of LLM-generated counterarguments in this domain remains an important empirical question.

Regarding prior experience using LLMs, previous research has shown that greater experience with AI is associated with more positive attitudes toward AI and greater approval of its use (Horowitz & Kahn, 2021). Although this research focused on general attitudes rather than the acceptance of AI-generated counterarguments, it suggests that participants with prior LLM experience may be more receptive to such counterarguments, whereas those without prior experience may respond more cautiously to an unfamiliar source.

**H3:** Individuals with lower confidence in their own judgments, higher perceived decision difficulty, prior experience using LLMs, and higher trust in AI will be more likely to accept LLM-generated counterarguments.

Although H3 includes several predictors, these predictors were not combined into a single composite index. Rather, each predictor was derived from a distinct line of previous research and was treated as a separate factor that may explain individual differences in judgment reversal following LLM-generated counterarguments.

# 4 Methods

## *4.1 Participants*

In the present study, younger adults were defined as those aged 18–30 years and older adults as those aged 65 years or older, consistent with common age-group conventions in cognitive aging research. Participants were recruited from a pool of individuals who had participated in the authors' previous studies on aging and cognition. The recruitment pool for older adults originally consisted of individuals randomly sampled from the Basic Resident Register, whereas younger adults were recruited through university bulletin boards and classroom announcements. Individuals in this pool were invited by telephone or email to participate in the present experiment. The final sample included 56 younger adults (age range: 19–28 years; M = 21.34 years, SD = 1.78; 35 women and 21 men) and 74 older adults (age range: 65–83 years; M = 74.23 years, SD = 5.06; 36 women and 38 men). All younger participants were enrolled in undergraduate or graduate programs. All older adults had attained at least a secondary level of education; 26 had completed high-school, 14 had completed two-year college, 31 had completed four-year university, and 3 had completed graduate school. Upon completion of the experiment, older adults received 10,000 yen and younger adults received 6,000 yen as compensation. The compensation for older adults included transportation costs, such as bus fares for two days, and their experimental session lasted approximately one hour longer than that of younger adults. Taking these factors into account, the compensation amounts were set so that the hourly compensation rate was approximately comparable across age groups.

Cognitive functioning was assessed using the Wechsler Adult Intelligence Scale–Fourth Edition (WAIS-IV; Wechsler, 2008). The WAIS-IV consists of 10 core subtests (see Section 4.4,

Cognitive Function), and the scores for each subtest in the younger and older groups are presented in Table 1. These 10 subtests are used to derive the Full Scale IQ. No participant had a Full Scale IQ below 70, and Full Scale IQ scores in the total sample ranged from 79 to 154. This indicates that no participant showed evidence of general intellectual impairment. Group means and standard deviations for the younger and older groups are reported in Table 1. In addition, no participant reported a history of dementia or a neurological or psychiatric disorder.

### *4.2 Design and Sample Size*

Age group was treated as a between-subjects factor, whereas moral dilemma (switch dilemma vs. footbridge dilemma) and timing of the AI counterargument (pre vs. post) were treated as within-subjects factors.

To comprehensively assess responsiveness to LLM-generated counterarguments, we examined two outcomes: judgment reversal and degree of judgment change. Judgment reversal captured whether participants shifted their judgment from one side of the scale to the other after the presentation of a counterargument—that is, whether their judgment shifted from appropriate to inappropriate, or vice versa—thereby reflecting a categorical change in moral judgment. In contrast, degree of judgment change captured the magnitude of change in participants' ratings, regardless of whether a complete reversal occurred, thereby reflecting more graded changes in judgment. Because these outcomes differed in their levels of measurement, different statistical models were used.

The primary dependent variable was judgment reversal, coded as a binary outcome: reversal = 1 and no reversal = 0. These data were analyzed using a generalized linear mixed model (GLMM) with a logistic link function. Age group, moral dilemma, and their interaction were entered as fixed effects, and participant ID was included as a random intercept. The adequacy of the sample size was evaluated after data collection through a post hoc sensitivity analysis based on the observed data using G*Power (Faul et al., 2007). In the present study, 130 participants responded to two moral dilemma conditions, yielding a total of 260 observations. However, because these observations were repeated measures nested within participants, they were not treated as independent. Instead, the design effect was taken into account based on the intraclass correlation coefficient (ICC = 0.227). As a result, the effective number of observations was estimated to be approximately 212, corresponding to an effective sample size of 106 participants. Furthermore, under a simplified set of assumptions in which the response probability in the reference category was set at .50, the significance level was set at .05, and statistical power was set at .80, the minimum detectable effect size was estimated to be approximately an odds ratio of 3.45. According to the criteria proposed by Chen et al. (2010), this value is close to a medium effect size (odds ratio = 3.47), indicating that the present study had sufficient statistical power to detect an effect of approximately medium magnitude.

To examine the degree of judgment change following the presentation of counterarguments, a mixed-design analysis of variance (ANOVA) was conducted with age group as the between-subjects factor and moral dilemma as the within-subjects factor. The dependent variable was the judgment change score, calculated by subtracting the pre-counterargument judgment rating from the post-counterargument judgment rating. A power analysis indicated that, assuming a statistical power of .80 and a medium effect size ($f = 0.25$), the required sample size was 98 participants. Because the present study included 130 participants, this criterion was satisfied, confirming that sufficient statistical power was secured for this analysis.

*4.3 Moral Dilemma Tasks*

The present study employed two moral dilemma tasks, the switch dilemma and the footbridge dilemma (Greene et al., 2004). Participants first completed a pre-counterargument phase in which they made their own judgments about each dilemma. In the pre-counterargument phase, the moral dilemma scenario was presented on a screen, and participants were instructed to read it aloud. They were then asked to judge whether it was appropriate to sacrifice one person in order to save five others, using a 6-point scale ranging from 1 (definitely inappropriate) to 6 (definitely appropriate). After making their judgment, participants were asked to rate their confidence in that judgment and the subjective difficulty of the judgment using the items "I am confident in this judgment" and "This judgment was difficult," respectively, on 5-point scales ranging from 1 (strongly disagree) to 5 (strongly agree). The order of presentation of the switch dilemma and the footbridge dilemma was counterbalanced across participants.

In the post-counterargument phase, participants who had indicated in the pre-counterargument phase that sacrificing one person to save five others was appropriate (i.e., ratings of 4, somewhat appropriate, to 6, definitely appropriate) were presented with a counterargument asserting that sacrificing one person to save five others was inappropriate. Conversely, participants who had indicated in the pre-counterargument phase that such a sacrifice was inappropriate (i.e., ratings of 1, definitely inappropriate, to 3, somewhat inappropriate) were presented with a counterargument asserting that sacrificing one person to save five others was appropriate.

The counterarguments were generated using GPT-4o in ChatGPT during the period between the September 3, 2024, update and the November 20, 2024, update. The counterarguments were designed to present logical arguments against participants' initial judgments from five distinct perspectives, rather than relying on emotional appeals or unsupported rejection. The full prompts used to generate the four counterarguments, together with the verbatim text of all four counterarguments as displayed to participants, are provided in the Supplementary Material. To fully standardize the counterargument content across participants, we did not have ChatGPT generate counterarguments in real time. Instead, we recorded videos showing the response text being generated sequentially within the ChatGPT interface and presented the relevant video on a monitor placed in front of each participant. This format was chosen both to standardize the counterargument content and to make clear that the counterarguments were AI-generated rather than produced by a human experimenter.

The experimenter verbally reminded each participant of the response they had provided to the dilemma in the pre-counterargument phase and then presented a counterargument against that response on the monitor. Participants were asked to read aloud the displayed counterargument and then to judge again whether it was appropriate to sacrifice one person to save five others. They then rated the statement, "The AI's counterargument was persuasive," on a 5-point scale ranging from 1 (strongly disagree) to 5 (strongly agree). For each participant, the dilemmas were presented in the same order in the counterargument phase as in the pre-counterargument phase. After completing the post-counterargument phase, participants were debriefed and informed that there were no correct answers to the two dilemma tasks. They were also told that two AI-generated counterarguments had been prepared for each dilemma, so that a counterargument could be presented regardless of whether they had initially judged the action to be appropriate or inappropriate. Participants were further informed that the AI had not determined that their judgment was incorrect.

### 4.4 Cognitive Function

Cognitive function was assessed using the Japanese version of the Wechsler Adult Intelligence Scale–Fourth Edition (WAIS-IV; Wechsler, 2008). The WAIS-IV consists of 10 subtests designed to measure four domains of cognitive functioning: the Verbal Comprehension Index (VCI), the Perceptual Reasoning Index (PRI), the Working Memory Index (WMI), and the Processing Speed Index (PSI). The VCI comprises the Similarities, Vocabulary, and Information subtests; the PRI comprises the Block Design, Matrix Reasoning, and Visual Puzzles subtests; the WMI comprises the Digit Span and Arithmetic subtests; and the PSI comprises the Symbol Search and Coding subtests. In addition to these subtest scores, the present study calculated the Full Scale IQ (FSIQ). Full Scale IQ, as reported in Table 1, is an age-normed standard score (normative M = 100, SD = 15) and therefore reflects performance relative to same-age peers, rather than absolute performance. In contrast, the subtest scores reported in the same table are raw scores and are not age-corrected. Thus, age-normed FSIQ scores and raw subtest scores should be interpreted differently when comparing younger and older adults.

### 4.5 Trust in AI

To assess trust in AI, the Trust Scale for Artificial Intelligence (Katase, 2021) was administered. This scale consists of two factors: Trust in the Social Benefit of AI, which includes five items such as “AI generally produces correct results,” and Distrust in the Fidelity of AI, which includes five items such as “AI sometimes deceives people.” Responses were recorded on a 7-point scale ranging from 1 (strongly disagree) to 7 (strongly agree).

### *4.6 Prior Experience Using LLMs*

Prior experience using LLMs was assessed with a single binary item asking participants whether they had previously used an LLM, such as ChatGPT. Responses were recorded as 0 = no and 1 = yes. This item was intended to capture participants' general familiarity with LLMs before the experiment, rather than their frequency of use or experience with moral-dilemma tasks specifically.

### *4.7 Procedure*

The experiment was conducted individually. After the purpose and content of the study had been explained, written informed consent was obtained from all participants. On the first day, the WAIS-IV was administered. On the second day, participants completed the decision-making tasks and were then asked to complete the measure assessing trust in AI.

## 5 Results

### *5.1 Age Differences in Study Variables*

Table 1 presents descriptive statistics and results of between-group comparisons by age group for cognitive function, trust in AI, subjective evaluations of dilemma judgments, and prior experience using LLMs. With regard to cognitive function, younger adults scored higher than older adults on all subtests except Vocabulary. Because verbal knowledge has been shown to remain relatively preserved in older adulthood, this pattern is consistent with the typical age-related profile of cognitive function. With respect to trust in AI, no significant age-group differences were observed. In addition, for both the switch dilemma and the footbridge dilemma, older adults reported significantly greater confidence in their judgments than younger adults, whereas no significant age-group differences were found in subjective decision difficulty.

Prior experience using LLMs also differed markedly by age group: 47 of the 56 younger adults (83.93%) reported prior experience using LLMs, whereas only 9 of the 74 older adults (12.16%) did so. Younger adults were significantly more likely than older adults to have prior experience using LLMs ($\chi^2(1) = 66.96$, $p < .001$).

### *5.2 Moral Dilemma Judgments in the Pre-counterargument Phase*

Figure 1 presents, for each age group, the proportions of appropriate and inappropriate responses in the switch dilemma and the footbridge dilemma before exposure to the LLM-generated counterarguments. Participants rated whether sacrificing one person to save five others was appropriate on a 6-point scale ranging from 1 (definitely inappropriate) to 6 (definitely appropriate). For the analysis, responses of 1 to 3 were classified as inappropriate, while responses of 4 to 6 were classified as appropriate.

In the switch dilemma, 64.61% of participants judged it appropriate to sacrifice one person to save five others, compared with 26.92% in the footbridge dilemma. Chi-square tests comparing the proportions of appropriate and inappropriate responses showed that, in both age groups, the proportion of participants who judged the action to be appropriate was significantly higher in the switch dilemma than in the footbridge dilemma (younger adults: $\chi^2(1) = 15.76$, $p < .001$; older adults: $\chi^2(1) = 21.76$, $p < .001$). In contrast, for neither dilemma did the proportions of appropriate and inappropriate responses differ significantly between age groups (switch dilemma: $\chi^2(1) = 1.09$, $p = .30$; footbridge dilemma: $\chi^2(1) = 1.36$, $p = .24$). These results indicate that the two dilemmas differed as anticipated prior to the experiment: the switch dilemma elicited more utilitarian judgments, whereas the footbridge dilemma elicited more deontological judgments.

### 5.3 Judgment Change Following LLM Counterarguments

First, to confirm that the perceived persuasiveness of the LLM-generated counterarguments did not differ across the two moral dilemmas, a paired-samples *t* test was conducted comparing persuasiveness ratings for the switch dilemma (M = 3.53, SD = 1.18) and the footbridge dilemma (M = 3.35, SD = 1.38). The results indicated no significant difference in the persuasiveness of the LLM counterarguments between the two dilemmas ($t(129) = 1.76$, $p = .08$, *Cohen's d* = .15, 95% CI for the mean difference [−0.02, 0.39]).

Figure 2 shows, for each age group, the proportions of participants in each dilemma who accepted the LLM-generated counterargument and reversed their judgment from appropriate to inappropriate or vice versa. In the switch dilemma, 23.21% of younger adults and 39.19% of older adults reversed their judgment. In the footbridge dilemma, 32.14% of younger adults and 40.54% of older adults reversed their judgment. To examine whether participants accepted the LLM-generated counterarguments and reversed their judgment, a generalized linear mixed model (GLMM) was conducted with judgment reversal (no reversal vs. reversal) as the dependent variable, age group (younger vs. older adults) and moral dilemma (switch vs. footbridge) as fixed effects, and participant as a random effect. The GLMM showed that neither the interaction between age group and moral dilemma ($F(1, 256) = 0.54$, $p = .46$) nor the fixed effect of moral dilemma ($F(1, 256) = 0.91$, $p = .34$) was significant. In contrast, the fixed effect of age group showed a marginal trend toward significance ($F(1, 256) = 3.08$, $p = .08$), suggesting that older adults tended to be more likely than younger adults to accept the LLM-generated counterarguments. Because the effect of moral dilemma on judgment reversal was not significant, H1, which predicted greater reversal in the switch dilemma than in the footbridge dilemma, was not supported.

Although the order of the two dilemmas was counterbalanced across participants, the influence of LLM-generated counterarguments may have differed depending on whether participants completed the switch dilemma or the footbridge dilemma first. Therefore, we conducted a supplementary GLMM that included dilemma order (switch-first vs. footbridge-first) and its interactions as fixed effects. The results showed that neither the fixed effect of dilemma order ($F(1, 252) = 0.52$, $p = .47$) nor any interaction involving dilemma order was significant ($Fs < 2.10$, $ps > .15$). Importantly, the marginal trend for the fixed effect of age group remained unchanged after including dilemma order in the model ($F(1, 252) = 3.19$, $p = .08$), indicating that judgment reversal was not systematically influenced by the order in which the dilemmas were presented.

Figure 3 shows the degree of judgment change following the LLM-generated counterarguments, calculated as the absolute difference between post-counterargument and pre-counterargument judgments. To examine the effects of age group and moral dilemma on the degree of judgment change, a mixed-design analysis of variance (ANOVA) was conducted, with age group as the between-subjects factor and moral dilemma as the within-subjects factor. The analysis revealed a significant main effect of age group ($F(1, 128) = 4.61$, $p = .03$, $\eta_p^2 = .04$) and a significant interaction between age group and moral dilemma ($F(1, 128) = 6.34$, $p = .01$, $\eta_p^2 = .05$). Because the interaction was significant, Bonferroni-corrected multiple comparisons were conducted. The results showed that, in the switch dilemma, older adults showed a significantly greater degree of judgment change than younger adults ($p < .001$), whereas no significant age-group difference was observed in the footbridge dilemma ($p$ = .62). In addition, among younger adults, the degree of judgment change was significantly greater in the footbridge dilemma than in the switch dilemma ($p$ = .05). Taken together with the GLMM results, the direction of the age effect was consistent with H2, which predicted that older adults would be more likely than younger adults to change their judgments in response to LLM-generated counterarguments.

To examine whether the degree of judgment change was affected by dilemma order, an additional mixed-design ANOVA was conducted with age group and dilemma order as between-subjects factors and moral dilemma as a within-subjects factor. The main effect of dilemma order was not significant ($F(1, 126) = 1.58$, $p = .21$, $\eta_p^2 = .01$), and no interactions involving dilemma order were significant ($Fs < 1.90$, $ps > .17$). Notably, the main effect of age group ($F(1, 126) = 4.61$, $p = .03$, $\eta_p^2 = .04$) and the interaction between age group and moral dilemma ($F(1, 126) = 6.33$, $p = .01$, $\eta_p^2 = .05$) remained significant after including dilemma order in the model. These results indicate that the degree of judgment change was not systematically influenced by the order in which the dilemmas were presented.

### *5.4 Judgment Reversal Following LLM Counterarguments and Cognitive Function*

To compare cognitive function between participants who reversed their judgments in response to the LLM-generated counterarguments and those who did not, separate one-way MANOVAs were conducted for each dilemma within each age group, with the WAIS-IV subtest scores as the dependent variables. Among younger adults, the multivariate effect was not significant for either the switch dilemma ($F(10, 45) = 0.84$, $p = .60$, $\eta_p^2 = .16$) or the footbridge dilemma ($F(10, 45) = 0.64$, $p = .78$, $\eta_p^2 = .12$), indicating no significant differences in cognitive functioning between those who reversed their judgment and those who did not. Among older adults, the multivariate effect was not significant in the switch dilemma ($F(10, 63) = 0.67$, $p = .75$, $\eta_p^2 = .10$), but it was significant in the footbridge dilemma ($F(10, 63) = 1.98$, $p = .05$, $\eta_p^2 = .24$). Follow-up ANOVAs (Table 2) showed that, with the exception of Coding, older adults who reversed their judgments had significantly lower cognitive function scores than those who did not. These findings provide partial support for H2, as lower cognitive functioning was associated with judgment reversal only among older adults in the footbridge dilemma.

### *5.5 Associations Between Judgment Reversal Following LLM Counterarguments, Trust in AI, and Subjective Evaluations of One's Judgment*

Logistic regression analyses were conducted separately for each dilemma (Table 3), with judgment reversal following exposure to the LLM-generated counterarguments as the dependent variable and age group, cognitive function (FSIQ), trust in AI, prior experience using LLMs, confidence in one's judgment, and perceived decision difficulty as predictors. The results showed that, in the switch dilemma, judgment reversal was more likely among older adults than younger adults, among those with lower confidence in their initial judgment, and among those who reported greater decision difficulty. In the footbridge dilemma, participants with lower cognitive function and those who perceived their initial judgment as more difficult were more likely to reverse their judgment in response to the counterargument. Importantly, because the predictors were measured on different response scales and the odds ratios reported in the present analyses were based on unstandardized coefficients, these odds ratios should not be interpreted as direct indicators of the relative strength of the predictors. With respect to H3, the prediction was only partially supported. Consistent with H3, lower confidence in one's initial judgment predicted greater judgment reversal in the switch dilemma, and higher perceived difficulty predicted greater judgment reversal in both dilemmas. Contrary to H3, however, neither trust in AI nor prior experience with LLMs significantly predicted judgment reversal in either dilemma.

# 6 Discussion

The purpose of this study was to examine whether LLM-generated counterarguments can change people's moral judgments, whether acceptance of such counterarguments differs between younger and older adults, and whether prior experience using LLMs and trust in AI influence judgment change in response to those counterarguments.

## *6.1 Judgment Change Following LLM Counterarguments*

The results showed that 32.31% of participants in the switch dilemma and 36.92% of those in the footbridge dilemma accepted the LLM-generated counterarguments and reversed their judgments. Previous research has suggested that people tend to exhibit egocentric advice discounting, that is, a tendency to incorporate only approximately 20% to 30% of others' advice into their own judgments (Bonaccio & Dalal, 2006). In this context, the complete judgment reversals observed in more than 30% of participants in the present study suggest that logically structured LLM-generated counterarguments can exert a non-trivial influence on moral judgment. Although the absence of a human-counterargument control condition calls for caution in attributing this influence specifically to the LLM, the present findings nevertheless demonstrate that counterarguments delivered through an LLM interface can produce substantial changes in moral judgment in a controlled experimental setting.

In the present study, we hypothesized, following the traditional dual-process account, that participants would be more likely to accept LLM-generated counterarguments and change their judgments in the switch dilemma than in the footbridge dilemma (H1). However, the proportion of participants who reversed their judgments did not differ significantly between the two dilemmas, and H1 was therefore not supported. This null finding does not necessarily contradict the dual-process framework, because the framework primarily concerns the formation of initial moral judgments rather than the revision of those judgments in response to external counterarguments. Moreover, recent developments in moral-dilemma research suggest that the distinction between the switch and footbridge dilemmas may not be reducible to a simple contrast between controlled cognition and emotion. For example, competing-intuitions accounts suggest that utilitarian responses to high-conflict moral dilemmas can arise intuitively, rather than necessarily through deliberate correction of an initial deontological response (Bago & De Neys, 2019; De Neys, 2023). Similarly, multi-process perspectives suggest that moral-dilemma judgments may reflect multiple distinguishable determinants, including sensitivity to consequences, sensitivity to moral norms, and a general preference for inaction versus action, rather than a single emotion-versus-cognition axis (Gawronski et al., 2017). From this perspective, the theoretical basis for predicting a simple dilemma-type difference in responsiveness to LLM-generated counterarguments becomes less straightforward. The present findings may therefore indicate that the difference between the switch and footbridge dilemmas observed at the stage of initial judgment formation does not directly generalize to judgment revision following LLM-generated counterarguments.

Two further, non-exclusive interpretations may also explain the absence of a dilemma-type difference. First, the evaluation of an externally supplied argument may involve a heuristic process that operates at a different stage than the formation of an initial moral judgment. Landes et al. (2026) proposed that, when evaluating LLM output, people may check for inconsistencies or obvious errors but, in the absence of a fatal flaw, defer to a response that seems sufficiently convincing or "good enough." Because such an evaluation-stage heuristic is complementary to, rather than competitive with, the dual-process account of judgment formation, a comparable reliance on this heuristic across the switch and footbridge dilemmas may have attenuated any dilemma-type difference in judgment reversal. Second, both dilemmas are subjective tasks with no objectively correct answer, and algorithm aversion tends to emerge especially in subjective, intuition-based tasks (Castelo et al., 2019). If participants showed a similar level of algorithm aversion across the two dilemmas, this too may have reduced any dilemma-type difference in the influence of the counterarguments.

### *6.2 Age Differences and Cognitive Offloading*

The present study examined the hypothesis that older adults would be more likely than younger adults to change their judgments in response to LLM-generated counterarguments (H2), especially among those with lower cognitive functioning. The results showed that older adults were more likely to reverse their judgments than younger adults in response to the LLM-generated counterarguments. In particular, in the switch dilemma, older adults exhibited a greater degree of judgment change than younger adults. These findings provide partial support for H2.

The GLMM and mixed-design ANOVA converged in suggesting that older adults were more strongly influenced by LLM-generated counterarguments than younger adults, although the age-group effect was only marginal in the GLMM ($p = .08$) and significant in the ANOVA. This pattern likely reflects what each analysis captures: the GLMM tests whether participants cross the categorical threshold for reversal, whereas the ANOVA quantifies the magnitude of judgment change as a continuous outcome. Participants who shifted substantially toward the counterargument without fully crossing the midpoint contribute to the latter but not the former. Age-related differences in responsiveness thus appear to emerge more clearly as graded shifts than as complete categorical reversals.

The finding that older adults were more strongly influenced by the LLM-generated counterarguments than younger adults may be understood in terms of cognitive offloading, defined as the use of external tools to reduce cognitive burden (Gerlich, 2025). Because aging is associated with declines in processing speed and working memory, older adults may be more likely to avoid engaging in complex thought in order to conserve cognitive resources (Hess, 2014). In the present study, the counterarguments generated by the LLM may have served as an external aid that reduced the cognitive burden associated with independently reasoning through morally difficult decisions. It is also noteworthy that, in the footbridge dilemma, which is thought to rely on emotional moral judgments that are relatively resistant to revision through logical reasoning (Haidt, 2001), older adults with lower cognitive functioning were more likely to reverse their judgments. This finding suggests that, even when the LLM presented a counterargument involving the emotionally aversive act of pushing one man to his death, older adults with lower cognitive functioning may have had greater difficulty maintaining or elaborating their initial judgment in the face of a logically structured counterargument. One possible interpretation is that these participants relied more heavily on the LLM-generated counterargument as an external cognitive resource and evaluated it through a “seems good enough” heuristic (Landes et al., 2026).

However, these findings should not be interpreted as definitive evidence that age-related cognitive decline alone caused greater responsiveness to AI-generated counterarguments. The younger and older groups differed substantially in prior LLM experience, and age was confounded with technology familiarity in the present design. Older adults' responses may therefore reflect not only cognitive offloading but also unfamiliarity with LLMs, novelty effects, or a greater tendency to perceive the technology as authoritative.

The larger age-related shift in the switch dilemma may also reflect the content of the counterarguments. Because many participants initially made utilitarian judgments in the switch dilemma, the opposing counterarguments often emphasized deontological considerations. Given that older adults tend to place greater weight on deontological concerns than younger adults, such counterarguments may have been especially influential among older adults (McNair et al., 2019).

### *6.3 Individual Factors Predicting Acceptance of LLM Counterarguments*

In the present study, we hypothesized that individuals with lower confidence in their own judgments, higher perceived difficulty, prior experience with LLMs, and higher trust in AI would be more likely to accept LLM-generated counterarguments and reverse their judgments (H3). This prediction was only partially supported. Consistent with H3, lower confidence in one's initial judgment and higher perceived difficulty predicted judgment reversal. Contrary to the trust-based component of H3, however, neither general trust in AI nor prior experience with LLMs predicted judgment reversal.

This dissociation between trust-related and judgment-related predictors represents an important contribution of the present analysis, because both the advice-taking and human–AI interaction literatures have treated source reliability, or trust in the source, as a primary determinant of whether external advice or algorithmic recommendations are accepted (Bonaccio & Dalal, 2006; Castelo et al., 2019; Yaniv, 2004). Our findings instead converge with, and extend, recent reports that reliance on LLM advice is not driven by general trust in AI (Ikeda, 2024) and that higher-order beliefs such as trust do not reliably translate into persuasive impact—a "disconnect between trust and persuasion" (Landes et al., 2026). Because this finding concerns general trust in AI rather than the perceived credibility of the specific counterargument, it qualifies accounts that emphasize general attitudes toward AI rather than contradicting advice-taking research as a whole. The present study further advances this work by moving beyond previous paradigms based on advice and recommendations to a setting in which the LLM directly counters participants' own prior judgments, and by testing this paradigm in a sample that includes older adults. In a domain where careful consideration of the specific features of a situation is emphasized (Castelo et al., 2019), participants appear to have evaluated the content of the counterargument itself rather than simply deferring to, or resisting, the LLM on the basis of their general attitudes toward AI.

One possible explanation for the null effect of prior LLM experience is that participants may have had little experience using LLMs in unusual tasks such as moral dilemmas. Consequently, prior experience with LLMs may not have been meaningfully related to the evaluation of LLM-generated counterarguments in the present task. Given that most older adults had no prior experience with LLMs, future research should further examine the role of prior LLM experience.

In contrast, subjective evaluations of one's own judgment, such as confidence and perceived difficulty, did influence whether participants subsequently reversed their judgments in response to the LLM-generated counterarguments. Research on advice taking has shown that the more confident individuals are in their own judgments, the more likely they are to prioritize those judgments over external advice (Yaniv, 2004). It has also been shown that task difficulty moderates advice use, such that advice tends to be weighted more heavily when tasks are more difficult (Gino & Moore, 2007). Similarly, a study on the persuasive effects of LLMs (Ikeda, 2024) reported that individuals with lower confidence in their own decisions are more susceptible to LLM advice, and that reliance on LLM-generated advice is not necessarily driven by trust in AI. Ikeda (2024) further suggested that, under conditions of high uncertainty, LLM advice may function as a reference point.

Taken together, these findings indicate that acceptance of LLM-generated counterarguments was shaped more by judgment-specific factors, namely lower confidence and greater perceived difficulty, than by general evaluations of AI, thereby addressing recent calls to examine the individual characteristics and psychological factors underlying AI persuasion (Hölbling et al., 2025).

## 7 Limitations

The present study has several methodological limitations. First, no human-counterargument control condition was included. Therefore, although the findings provide evidence that logically structured counterarguments delivered through an LLM interface can exert a meaningful influence on moral judgment, we cannot determine whether the observed judgment reversal rate of more than 30% reflects an effect unique to LLM-generated arguments or whether a similar effect would have occurred in response to any logically coherent counterargument.

Second, the two age groups differed substantially in prior LLM experience. Thus, the present design cannot fully disentangle age-related cognitive influences from cohort-level novelty effects or technology familiarity. In the logistic regression analyses, age group remained a significant predictor in the switch dilemma when entered together with prior LLM experience, whereas prior LLM experience was not significant in either dilemma (Table 3), providing some evidence against a purely novelty-driven account. Nevertheless, the extreme imbalance between age groups and the binary measurement of prior use limit the strength of this conclusion. Future research should include middle-aged adults or recruit samples matched on LLM experience.

Third, the older adult sample was cognitively high-functioning, with a mean FSIQ above the population norm. Therefore, the present findings may underestimate the extent to which lower cognitive functioning is associated with responsiveness to LLM-generated counterarguments in more representative older adult populations.

Fourth, although we use the term susceptibility to describe participants' tendency to revise their judgments following LLM-generated counterarguments, the present design cannot determine whether such revisions reflect adaptive belief updating, cognitive offloading, conformity, or undue persuasion. Because the dilemmas had no objectively correct answers or independent criterion of judgment quality, these judgment changes should not be interpreted as inherently maladaptive.

Fifth, the study employed an unusual moral dilemma as the decision-making task. Specifically, the trolley problem was used to examine whether acceptance of AI-generated counterarguments differs depending on whether judgments rely more heavily on cognitive control processes or on emotional and automatic processes. However, the extent to which these findings generalize to more naturalistic settings remains unclear. Further research is needed using tasks that more closely reflect everyday situations in which LLMs are likely to be used. Additionally, the study did not include direct bidirectional interaction between participants and the LLM. Because requiring older adults to operate the LLM directly was considered difficult, participants were shown only a video of ChatGPT-4o generating responses, and no real-time interaction took place between participants and the LLM. However, previous research has suggested that dialogic interaction may enhance persuasiveness (Hölbling et al., 2025; Salvi et al., 2025). Therefore, the present findings may underestimate the persuasive power of LLMs.

Sixth, the perspectives used to construct the LLM-generated counterarguments were not fully held constant across dilemmas or response directions. Some perspectives may have been more persuasive than others and may therefore have influenced judgment reversals. Although perceived persuasiveness did not significantly differ between the switch and footbridge dilemmas, the present design cannot isolate the effect of each perspective. Future research should systematically control argumentative perspectives when examining judgment reversal.

## 8 **Conclusion**

The present study demonstrated that, in moral dilemmas with no objectively correct answer, LLM-generated counterarguments can substantially influence and even reverse human judgment. Overall, the findings highlight the double-edged role of LLMs in human decision-making. For older adults, LLMs may function as supportive tools for cognitive offloading, yet the same mechanism may also increase their susceptibility to persuasive influence, particularly among those with lower cognitive functioning. Across age groups, low decision confidence and high perceived difficulty were associated with greater judgment reversal, suggesting that judgment-specific uncertainty may be more important than general trust in AI in predicting responsiveness to LLM-generated counterarguments. Because trolley dilemmas are abstract philosophical scenarios, future research should examine whether these findings generalize to everyday moral decision-making contexts and to more interactive forms of LLM use.

**Declaration of Interests**

The authors declare that they have no known competing financial interests or personal relationships that could have appeared to influence the work reported in this paper.

**Declaration of generative AI and AI-assisted technologies in the manuscript preparation process.**

During the preparation of this work the authors used ChatGPT (OpenAI) in order to assist with grammatical refinement, because the authors are not native speakers of English. After using this service, the authors reviewed and edited the content as needed and take full responsibility for the content of the published article.

**Open practices statement**

This study was not preregistered.

## Data Availability

The data supporting the findings of this study are available on Figshare via the following link: https://doi.org/10.6084/m9.figshare.32665356.

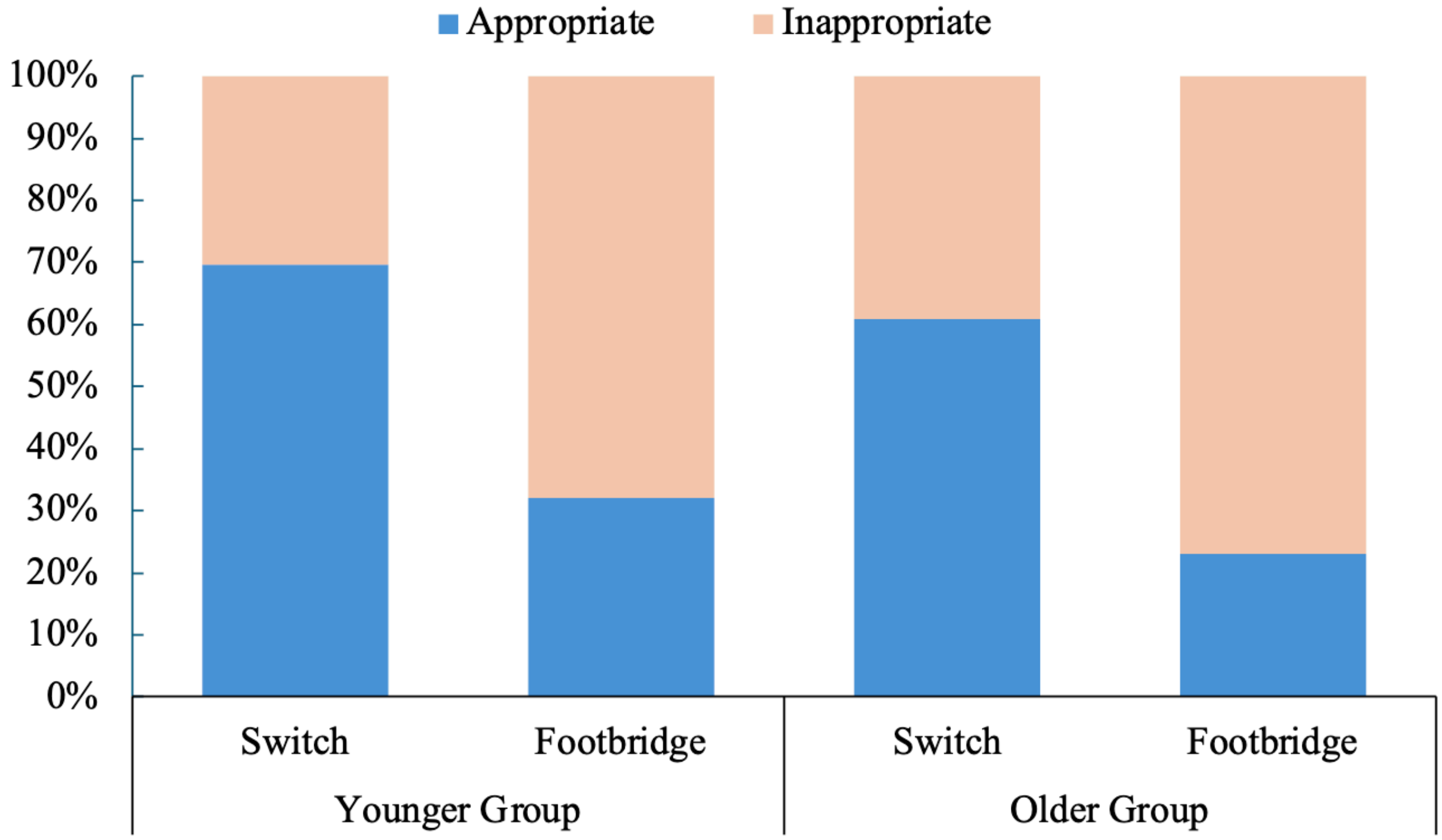

**Fig. 1.** Judgments in the trolley dilemma.

Note. Switch = switch dilemma; Footbridge = footbridge dilemma. Appropriate = judged it appropriate to sacrifice one person to save five others; Inappropriate = judged it inappropriate to do so. In the pre-counterargument phase, appropriate judgments were more frequent for the switch dilemma, whereas inappropriate judgments were more frequent for the footbridge dilemma. No significant age-group differences were observed for either dilemma.

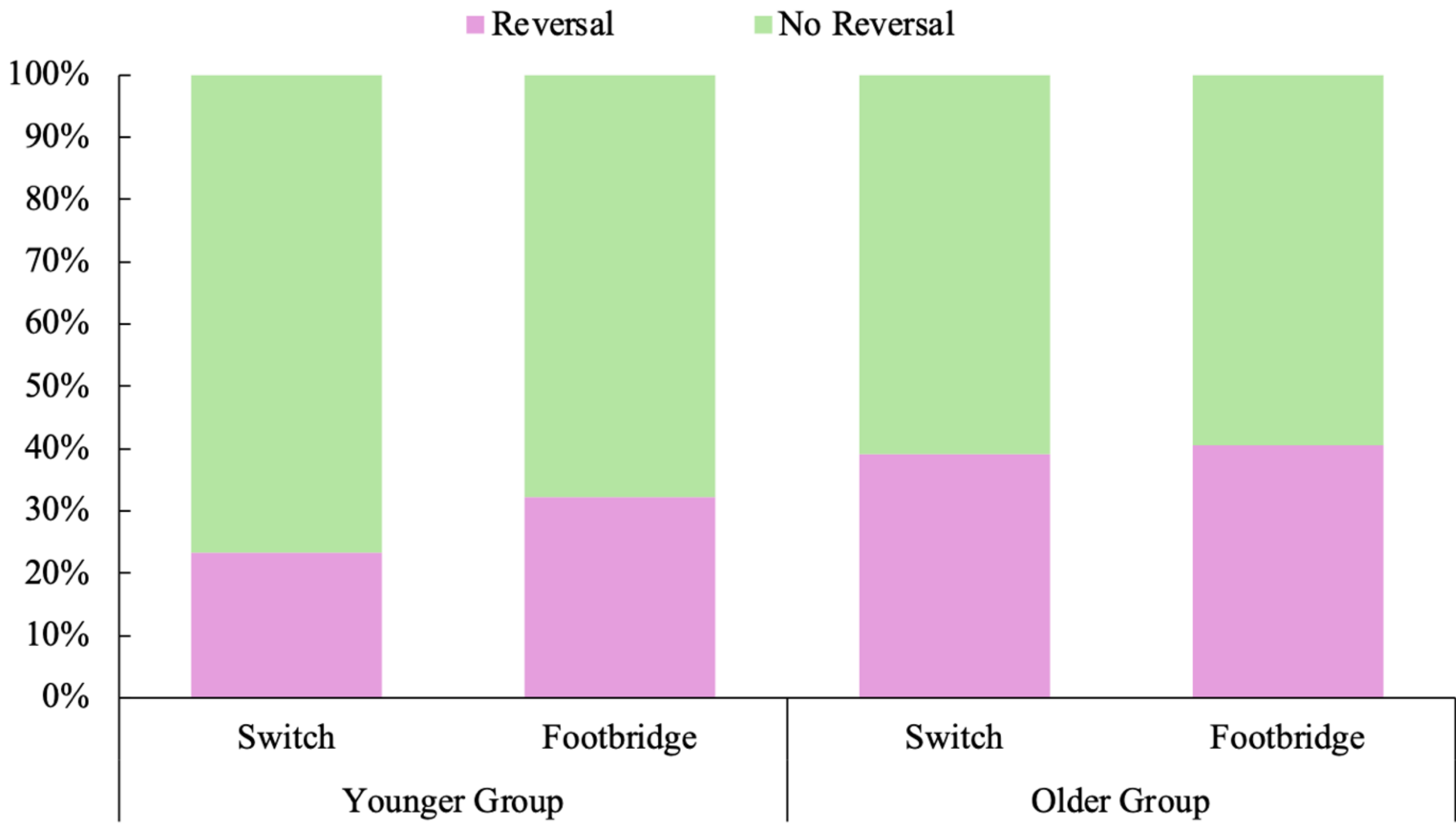

**Fig. 2.** Proportion of judgment reversal following LLM-generated counterarguments.

Note. Switch = switch dilemma; Footbridge = footbridge dilemma. Reversal = a participant's judgment shifted from one side of the appropriate/inappropriate midpoint of the 6-point scale to the opposite side after the LLM-generated counterargument. No reversal = the participant's judgment remained on the same side of the midpoint after the LLM-generated counterargument. The proportion of reversals tended to be higher in the older group than in the younger group ($p$ = .08).

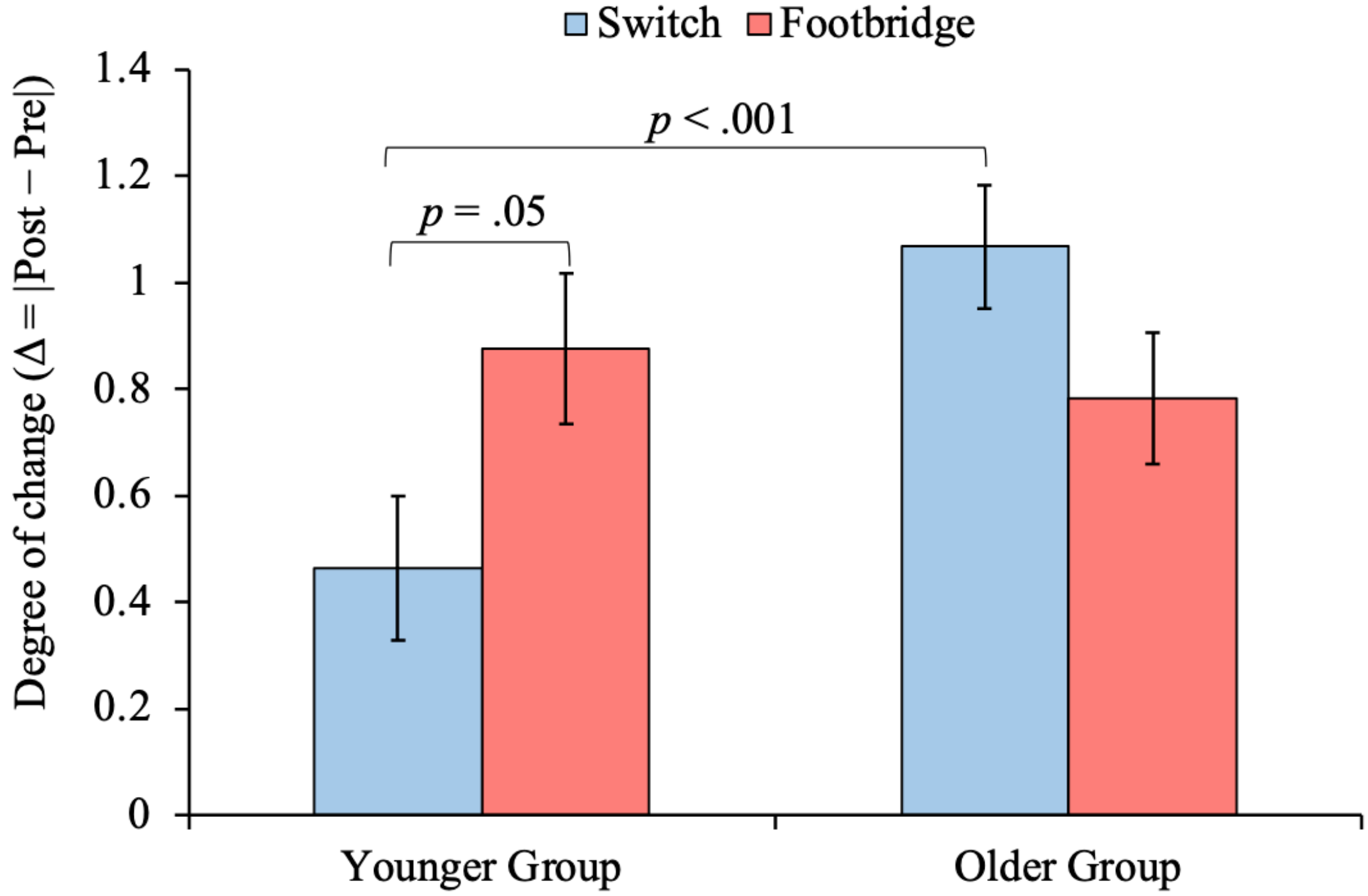

**Fig. 3.** Degree of judgment change following the LLM counterargument.

Note: Error bars indicate standard errors. Switch = switch dilemma; Footbridge = footbridge dilemma.

**Table 1**

*Age differences in cognitive function, trust in AI, subjective evaluations of dilemma judgments, and prior experience using LLMs.*

| **Variable** | **Younger** (n = 56) | **Older** (n = 74) | ***t*** | ***p*** | **Cohen's *d*** | **95% CI for the difference** |
|---|---|---|---|---|---|---|
| | M (SD) | M (SD) | | | | [LL, UL] |
| **Cognitive Function** | | | | | | |
| Full Scale IQ | 118.84 (8.14) | 115.84 (14.98) | 1.36 | .18 | 0.24 | [-1.38, 7.39] |
| *Verbal Comprehension* | | | | | | |
| Similarities | 26.95 (3.75) | 23.46 (5.17) | 4.27 | < .001 | 0.75 | [1.87, 5.10] |
| Vocabulary | 35.88 (6.98) | 34.26 (9.44) | 1.08 | .28 | 0.19 | [-1.35, 4.59] |
| Information | 18.21 (3.15) | 16.03 (4.83) | 2.95 | < .001 | 0.52 | [0.72, 3.66] |
| *Perceptual Reasoning* | | | | | | |
| Block Design | 58.96 (5.73) | 38.11 (9.34) | 14.73 | < .001 | 2.59 | [18.05, 23.66] |
| Matrix Reasoning | 21.05 (2.82) | 14.91 (4.22) | 9.41 | < .001 | 1.66 | [4.86, 7.44] |
| Visual Puzzles | 20.09 (2.68) | 13.04 (4.08) | 11.22 | < .001 | 1.97 | [5.81, 8.29] |
| *Working Memory* | | | | | | |
| Digit Span | 34.79 (5.87) | 25.20 (4.60) | 10.44 | < .001 | 1.84 | [ 7.77, 11.40] |
| Arithmetic | 18.79 (1.80) | 14.99 (3.51) | 7.40 | < .001 | 1.30 | [ 2.78, 4.82] |
| *Processing Speed* | | | | | | |
| Coding | 103.80 (14.09) | 67.24 (16.32) | 13.40 | < .001 | 2.36 | [31.16, 41.96] |
| Symbol Search | 45.79 (6.97) | 29.34 (7.12) | 13.16 | < .001 | 2.32 | [13.98, 18.92] |
| **Trust in Artificial Intelligence** | | | | | | |
| Trust in the social benefit of AI | 23.52 (3.89) | 23.07 (4.22) | 0.62 | .54 | 0.11 | [-0.98, 1.88] |

| Variable | Younger (n = 56) | Older (n = 74) | $t$ | $p$ | Cohen's $d$ | 95% CI for the difference |
|---|---|---|---|---|---|---|
| | M (SD) | M (SD) | | | | [LL, UL] |
| Distrust in the fidelity of AI | 20.95 (5.35) | 19.84 (4.04) | 1.35 | .18 | 0.24 | [-0.52, 2.74] |
| **Subjective evaluations of the trolley dilemma** | | | | | | |
| *Switch dilemma* | | | | | | |
| Judgment confidence | 2.86 (1.26) | 3.70 (1.29) | -3.74 | < .001 | -0.66 | [-1.29, -0.40] |
| Judgment difficulty | 4.14 (1.35) | 4.01 (1.38) | 0.53 | .59 | 0.10 | [-0.35, 0.61] |
| *Footbridge dilemma* | | | | | | |
| Judgment confidence | 3.05 (1.15) | 3.57 (1.39) | -2.25 | .03 | -0.40 | [-0.97, -0.06] |
| Judgment difficulty | 3.95 (1.42) | 3.77 (1.53) | 0.67 | .50 | 0.12 | [-0.34, 0.70] |
| **Prior Experience Using LLMs** | | | | | | |
| Prior LLM experience, *n* (%) | 47 (83.93%) | 9 (12.16%) | 66.96[a] | < .001 | 0.72[b] | — |

Note. Prior experience using LLMs is a binary variable and is reported as n (%); the age-group difference was tested with a chi-square test. [a] Value in the *t* column is the $\chi^2(1)$ statistic. [b] Effect size is *Cramér's V*, not *Cohen's d.*

**Table 2**

*Cognitive function in older adults who did and did not reverse their judgments in response to the AI-generated counterargument in the footbridge dilemma.*

| **WAIS-IV subtest** | **No reversal** (n = 44) | **Reversal** (n = 30) | ***F*** | ***p*** | $\eta_p^2$ | **95% CI** |
|---|---|---|---|---|---|---|
| | M (SD) | M (SD) | | | | [LL, UL] |
| *Verbal Comprehension* | | | | | | |
| Similarities | 25.00 (5.03) | 21.20 (4.57) | 10.96 | < .001 | 0.13 | [1.51, 6.09] |
| Vocabulary | 36.25 (8.12) | 31.33 (10.57) | 5.11 | .03 | 0.07 | [0.58, 9.25] |
| Information | 17.14 (4.46) | 14.40 (4.95) | 6.13 | .02 | 0.08 | [0.53, 4.94] |
| *Perceptual Reasoning* | | | | | | |
| Block Design | 40.14 (9.80) | 35.13 (7.88) | 5.42 | .02 | 0.07 | [0.72, 9.29] |
| Matrix Reasoning | 16.14 (4.20) | 13.10 (3.61) | 10.41 | < .001 | 0.13 | [1.16, 4.91] |
| Visual Puzzles | 14.16 (3.82) | 11.40 (3.95) | 9.04 | < .001 | 0.11 | [0.93, 4.59] |
| *Working Memory* | | | | | | |
| Digit Span | 26.20 (4.67) | 23.73 (4.13) | 5.47 | .02 | 0.07 | [0.37, 4.58] |
| Arithmetic | 15.66 (3.45) | 14.00 (3.41) | 4.16 | .05 | 0.06 | [0.04, 3.28] |
| *Processing Speed* | | | | | | |
| Coding | 70.05 (17.36) | 63.13 (13.94) | 3.30 | .07 | 0.04 | [-0.67, 14.50] |
| Symbol Search | 30.91 (7.01) | 27.03 (6.75) | 5.62 | .02 | 0.07 | [0.62, 7.14] |

Note. Reversal = a participant's judgment shifted from one side of the appropriate/inappropriate midpoint of the 6-point scale to the opposite side after the LLM-generated counterargument. No reversal = the participant's judgment remained on the same side of the midpoint after the LLM-generated counterargument.

**Table 3**

*Logistic regression analysis predicting judgment reversal following LLM-generated counterarguments.*

| **Predictor** | **Judgment reversal in the Switch dilemma** (0 = no reversal, 1 = reversal) | | | | **Judgment reversal in the Footbridge dilemma** (0 = no reversal, 1 = reversal) | | | |
|---|---|---|---|---|---|---|---|---|
| | ***B*** | **SE** | **OR [95% CI]** | ***p*** | ***B*** | **SE** | **OR [95% CI]** | ***p*** |
| Age Group (1 = Younger, 2 = Older) | 1.53 | 0.62 | 4.62 [1.38, 15.47] | .01** | 0.20 | 0.60 | 1.23 [0.38, 3.95] | .73 |
| Cognitive function (FSIQ) | -0.01 | 0.02 | 0.99 [0.95, 1.02] | .38 | -0.06 | 0.02 | 0.94 [0.91, 0.98] | .002** |
| Trust in the Social Benefit of AI | 0.06 | 0.06 | 1.07 [0.96, 1.19] | .25 | -0.01 | 0.05 | 0.99 [0.89, 1.10] | .85 |
| Distrust in the Fidelity of AI | -0.03 | 0.05 | 0.97 [0.88, 1.07] | .50 | 0.01 | 0.05 | 1.01 [0.92, 1.11] | .80 |
| Prior experience using LLM (0 = no, 1 = yes) | 0.28 | 0.58 | 1.32 [0.42, 4.09] | .63 | -0.29 | 0.60 | 0.75 [0.23, 2.44] | .63 |
| Judgment confidence (pre) | -0.56 | 0.18 | 0.57 [0.41, 0.81] | .002** | -0.30 | 0.16 | 0.74 [0.54, 1.01] | .06 |
| Judgment difficulty (pre) | 0.36 | 0.18 | 1.44 [1.01, 2.04] | .04* | 0.35 | 0.16 | 1.41 [1.04, 1.92] | .03* |
| model fit | $\chi^2(7) = 23.20$, $p = .002$**, $R^2 = .14$ | | | | $\chi^2(7) = 23.93$, $p = .001$***, $R^2 = .14$ | | | |

Note. Reversal = a participant's judgment shifted from one side of the appropriate/inappropriate midpoint of the 6-point scale to the opposite side after the LLM-generated counterargument. No reversal = the participant's judgment remained on the same side of the midpoint after the LLM-generated counterargument. OR = odds ratio. * $p < .05$, ** $p < .01$, ***$p < .001$.